\documentclass[preprint,showpacs,preprintnumbers,amsmath,amssymb]{revtex4}

\draft

\begin{document}
\title{Compositeness Effects in the Bose-Einstein Condensation}
\author{S.S. Avancini and J. R. Marinelli}
\address{Departamento de F\'{\i}sica - CFM, Universidade Federal de Santa\\
Catarina \\
C. P. 476 - 88040-900 Florian\'opolis - SC - Brazil}
\author{G. Krein}
\address{Instituto de F\'{\i}sica Te\'orica - Universidade Estadual Paulista \\
Rua Pamplona, 145 - 01405-900 S\~ao Paulo-SP - Brazil}

\begin{abstract}

Small deviations from purely bosonic behavior of trapped atomic
Bose-Einstein condensates are investigated with the help of the
quon algebra, which interpolates between bosonic and fermionic
statistics. A previously developed formalism is employed to obtain
a generalized version of the Gross-Pitaeviskii equation. Two
extreme situations are considered, the collapse of the condensate
for attractive forces and the depletion of the amount of condensed
atoms with repulsive forces. Experimental discrepancies observed
in the parameters governing the collapse and the depletion of the
condensates can be accounted for by universal fittings of the
deformation parameter for each case.

\vspace{0.25cm} PACS numbers:  03.75.Fi;05.30.Jp;03.65.Fd

\end{abstract}
\maketitle

\section{Introduction}

In many physical problems one has to deal with a large number of
identical particles that are not of a fundamental character but which are
known to be composed by a bound system of several fermions. Examples include
systems of identical atoms, molecules or nuclei. If the number of
``fundamental" fermions contained in the composite particle is odd, it is a
fermion, otherwise it is a boson. In many situations the internal structure
of the composite particle can be ignored and the system as a whole treated
as a collection of interacting or noninteracting point-like particles. This
is the case for instance, in the theories of Bose-Einstein condensation
(BEC) of trapped bosonic gases~\cite{Bec}. Another example is provided by
electron-hole bosonic states in semiconductors, called excitons~\cite
{Comb-epl}. In such systems, the bosons have internal structure and finite
size, i.e., they are composite bosonic particles. The rational for
neglecting the internal structure of the atoms in atomic BEC is that one is
dealing with a very dilute system in the trap. The low density regime makes
it very improbable that the internal structures of the atoms overlap in the
trap, since the average distance between atoms in typical condensates is
several times the size of an atom. For excitons, the situation is not as
favorable as in atomic BEC and effects of the internal structure of the
bosons might play an important role~\cite{Comb-epl}.

A departure from the purely bosonic behavior of the atoms in a trap might
occur in situations that the central density of the condensate grows beyond
a critical value. This happens, for example, when interatomic attractive
forces tend to push the atoms to the center of the trap and the zero-point
kinetic energy is no longer able to stabilize the system. The collapse is
expected to occur when the number $N$ of atoms exceeds a critical value,
$N_{\rm cr}$, leading to an interaction energy larger than the kinetic
energy.

The aim of the present paper is to set up a framework to evaluate the
departure from the purely bosonic behavior of a BEC of composite particles.
A complete theory aimed at such a task should include all the possible
degrees of freedom for the constituent particles, which is in general highly
prohibitive from the computational point of view. In the present paper, we
use a phenomenological approach, making use of the concept of {\em quons}~
\cite{GreenPRD}. Quons are particles that are neither bosons, nor fermions,
and the quon creation and annihilation operators obey a particular algebra
that interpolates between Fermi e Bose algebras. The quon algebra is in fact
a deformation of the Fermi and Bose algebras, and is such that when a
parameter ($q$) runs from $-1$ and $+1$, it interpolates between the Fermi
and Bose algebras.

Recently, a systematic way to build a many-body quon state has
been discussed~\cite{GreenPRD,SidRic} and a general formula for a
normalized many-quon symmetric state~\cite{SidRic} has been found.
The developed formalism allows to apply the quon algebra to
describe physical systems presenting deviations from an idealized
situation,  in the same fashion of the more usual Quantum
Algebras~\cite {Kibler}. An example is~\cite{SidRic} the three
dimensional quonic  harmonic oscillator which describes in a
effective way anharmonic effects. The quonic deformation is also
effective for incorporating correlations in many-body systems.
Applications involving the antisymmetric subspace (fermion-like
particles) can be also considered and are now under way.

The use of the symmetric subspace, and $q$ close enough to $+1$, allows to
describe in a very natural way the departure from a purely bosonic behavior
of systems of composite bosons. In order to apply such ideas to the BEC in
trapped gases, we derive in Section~\ref{sec:qGP} a quonic version of the
Gross-Pitaeviskii (GP) equation, which we denote qGP. In the limit of $q=1$,
the qGP equation reduces to the usual GP equation, widely used in the
literature~\cite{Bec}. Initially, in Section~\ref{sec:Comp-q}, we present
a brief discussion on the
description of a composite boson built from two non-identical fermions and
provide a connection with the quon algebra. This will allow a first estimate
of the deformation parameter $q$ in terms of parameters of the system~\cite
{Comb-epl,us}. In Section~\ref{sec:numer} we present numerical results, and
compare with observables related to trapped bosonic atoms. Conclusions and
future perspectives are presented in Section~\ref{sec:concl}.


\section{Relation between composite particles and quons}

\label{sec:Comp-q}

\subsection{Composite particles}

Let us consider a composite boson state with quantum number $\alpha$ as a
bound state of two distinct fermions
\begin{equation}
A^{\dagger}_{\alpha}|0\rangle=\sum_{\mu \nu} \Phi_{\alpha}^{\mu
\nu}a^{\dag}_{\mu}b^{\dag}_{\nu}|0\rangle\;,  \label{Adag}
\end{equation}
where $\Phi_{\alpha}^{\mu \nu}$ is the Fock-space bound-state amplitude,
$a^{\dag}_{\mu}$ and $b^{\dag}_{\mu}$ are the fermion creation operators, and
$|0\rangle$ is the no-particle state (vacuum). The quantum number $\alpha$
stands for the center of mass momentum, the internal energy, the spin, and
other internal degrees of freedom of the composite boson. For instance, the
composite boson could be the hydrogen atom, where $a^{\dag}_{\mu}$ and $%
b^{\dag}_{\mu}$ create an electron and a proton respectively. The $\mu$ and $%
\nu$ stand for the space and internal quantum numbers of the constituent
fermions. The sum over $\mu$ and $\nu$ is to be understood as a sum over
discrete quantum numbers and an integral over continuous variables.

The fermion creation and annihilation operators satisfy canonical
anti-commutation relations:
\begin{eqnarray}
&&\{a_{\mu}, a_{\nu}\}=\{a^{\dag}_{\mu}, a^{\dag}_{\nu}\}=0 \hspace{2.0cm}
\{b_{\mu}, b_{\nu}\}=\{b^{\dag}_{\mu}, b^{\dag}_{\nu}\}=0 \;,  \nonumber \\
&&\{a_{\mu}, a^{\dag}_{\nu}\}=\delta_{\mu,\nu} \hspace{3.65cm} \{b_{\mu},
b^{\dag}_{\nu}\}=\delta_{\mu,\nu}\;.  \label{fcomm}
\end{eqnarray}
It is convenient to work with normalized amplitudes $\Phi_{\alpha}^{\mu \nu}$,
such that
\begin{equation}
\langle\alpha|\beta\rangle=\delta_{\alpha,\beta}\;,  \label{1norm}
\end{equation}
and therefore
\begin{equation}
\sum_{\mu \nu}\Phi_{\alpha}^{\mu \nu *}\Phi_{\beta}^{\mu \nu}=
\delta_{\alpha,\beta}\;.  \label{Phinorm}
\end{equation}

Using the fermion anticommutation relations of Eq.~(\ref{fcomm}) and the
Fock-space amplitude normalization Eq.~(\ref{Phinorm}), one can easily show
that the composite boson operators satisfy the following commutation
relations:
\begin{eqnarray}
&&[A_{\alpha}, A_{\beta}]=[A^{\dag}_{\alpha}, A^{\dag}_{\beta}]=0\;,
\label{AAcomm}
\end{eqnarray}

\begin{eqnarray}
&&[A_{\alpha}, A^{\dag}_{\beta}]=\delta_{\alpha,\beta}- \Delta_{\alpha
\beta}\;,  \label{Acomm}
\end{eqnarray}
where $\Delta_{\alpha \beta}$ is given by
\begin{equation}
\Delta_{\alpha \beta}=\sum_{\mu \nu} \Phi^{\mu \nu *}_{\alpha}
\left(\sum_{\mu^{\prime}} \Phi^{\mu^{\prime}\nu}_{\beta}
a^{\dag}_{\mu^{\prime}}a_{\mu} + \sum_{\nu^{\prime}} \Phi^{\mu
\nu^{\prime}}_{\beta} b^{\dag}_{\nu^{\prime}}b_{\nu}\right)\;.  \label{Delta}
\end{equation}
One can also easily show the following commutation relations:
\begin{equation}
[a_{\mu},
A_{\alpha}^{\dag}]=\sum_{\mu^{\prime}\nu^{\prime}}\delta_{\mu,\mu^{\prime}}
\Phi_{\alpha}^{\mu^{\prime}\nu^{\prime}}b_{\nu^{\prime}}^{\dag} \hspace{2.0cm%
} [b_{\nu},
A_{\alpha}^{\dag}]=-\sum_{\mu^{\prime}\nu^{\prime}}\delta_{\nu,\nu^{\prime}}
\Phi_{\alpha}^{\mu^{\prime}\nu^{\prime}}a_{\mu^{\prime}}^{\dag}\;.
\label{indep}
\end{equation}

The composite nature of the bosons is evident from the presence of $%
\Delta_{\alpha \beta}$, which is a sort of ``deformation" of the canonical
boson algebra. The effect of this term becomes unimportant in the infinite
tight binding limit, i.e. in the limit of point-like bosons. Eq.~(\ref{indep}%
) shows the lack of kinematical independence of the microscopic operators $%
a_{\mu}$ and $b_{\nu}$ from the macroscopic ones $A_{\alpha}^{\prime}$s.


\subsection{The quon algebra}


The quon algebra is defined by the deformed commutation relation
\begin{eqnarray}
A_{\alpha}A^{\dag}_{\beta}-q A^{\dag}_{\beta}A_{\alpha}=
\delta_{\alpha,\beta} ,  \label{qAcomm}
\end{eqnarray}
where $q$ is the deformation parameter of the algebra; $A_{\alpha}$
annihilates the vacuum

\begin{equation}
A_{\alpha}|0\rangle=0\;.  \label{annvac}
\end{equation}
The quon algebra interpolates between the Fermi and Bose algebras as the
parameter $q$ varies from $-1$ to $+1$. Polynomials in the creation
operators acting on the vacuum form a Fock-like space of vectors, i.e., the
quonic Fock space. In Ref.~\cite{Zagier} it was shown that the squared norm
of all vectors in the quonic Fock space remains positive definite, provided
that $q$ ranges from $-1$ to $+1$.  \noindent Note that no commutation
relation can be imposed on $A^{\dag}_{\alpha}$,$A^{\dag}_{\beta}$ and $%
A_{\alpha}$,$A_{\beta}$. However, as remarked by Greenberg\cite{GreenPRD},
similarly to the case of normal Bose commutation relations, no such rule is
needed for practical evaluation of expectation values of polynomials in $%
A_{\alpha}$ and $A^{\dag}_{\alpha}$ when Eq.~(\ref{annvac}) holds. Such
matrix elements can be evaluated with the repeated use of Eq.~(\ref{qAcomm})
solely; annihilation operators are moved to the right using Eq.~(\ref{qAcomm}%
) until they annihilate the vacuum or creation operators are moved to the
left using the adjoint of Eq.~(\ref{qAcomm}) until they annihilate the
vacuum.

Since there is no commutation relations between $A^{\dag}_{\alpha}$,$%
A^{\dag}_{\beta}$ and $A_{\alpha}$,$A_{\beta}$ all $n!$ states given by the
order permutations of the creation operators acting on the vacuum (and
carrying different quantum numbers) are linearly independent. This is in
sharp contrast to the usual bosonic or fermionic algebras and important
differences between the quonic and usual Fock space appears due to this
property. As an example, we consider a two quons system occupying the
arbitrary states $1$ and $2$
\begin{eqnarray}
|1\rangle=A^{\dag}_{1}A^{\dag}_{2}|0\rangle
~~~~|2\rangle=A^{\dag}_{2}A^{\dag}_{1}|0\rangle~,
\end{eqnarray}
which are not independent states and the overlap matrix is
\begin{eqnarray}
\begin{pmatrix}
 \langle1|1\rangle & \langle1|2\rangle \\
 \langle2|1\rangle & \langle2|2\rangle
\end{pmatrix}=
 \begin{pmatrix}
  1 & q \\
  q & 1
  \end{pmatrix} .
\end{eqnarray}
One can show that the basis that diagonalizes the overlap matrix is
\begin{eqnarray}
|S\rangle &=& {\frac{1}{{\sqrt{2}\sqrt{1-q}}}}(A^{\dag}_1A^{\dag}_2 +
A^{\dag}_2A^{\dag}_1 )\mid 0 ,\rangle \\
|A\rangle &=& {\frac{1}{{\sqrt{2}\sqrt{1+q}}}}(A^{\dag}_1A^{\dag}_2 -
A^{\dag}_2A^{\dag}_1 )\mid 0\rangle ,
\end{eqnarray}
where $|S\rangle$ is a symmetric state and $|A\rangle$ is an antisymmetric
state. Analogously for a three quons system occupying three distinct states $%
1,2,3$, the non-orthogonal basis will contain the six independent states $%
A^{\dag}_i A^{\dag}_j A^{\dag}_k |0 \rangle $, where the indexes $i,j,k$ are
determined by the 3! permutations of $1,2,3$. The orthonormal basis may be
obtained classifying these states according to the irreducible
representations of the permutation group of three elements, $S_3$.

Besides the well known symmetric and antisymmetric states, there are four
more exotic mixed symmetric states. To shorten the corresponding expressions
we adopt the convention $A_i^{\dagger}A_j^{\dagger }A_k^{\dagger,
}|0\rangle\equiv |ijk\rangle$ and can write:
states:
\begin{eqnarray}
|S\rangle &=& \frac{1}{\sqrt{1+2q^2+2q+q^3}} \frac {1}{\sqrt{6}}\left[%
|ijk\rangle+|jik\rangle+|ikj\rangle +|jki\rangle+|kij\rangle+|kji\rangle%
\right] ,  \label{S} \\
|A\rangle &=& \frac 1{\sqrt{1+2q^2-2q-q^3}} \frac 1{\sqrt{6}}\left[%
|ijk\rangle-|jik\rangle-|ikj\rangle+|jki\rangle +|kij\rangle-|kji\rangle%
\right] , \label{A} \\
|MS1\rangle &=& \frac 1{\sqrt{1-q^2+q-q^3}}\frac 1{\sqrt{12}}%
[|ijk\rangle-|jik\rangle+2|ikj\rangle+|jki\rangle-2|kij\rangle-|kji\rangle] ,
\label{MS1} \\
|MS2\rangle &=& \frac 1{\sqrt{1-q^2+q-q^3}}\frac 12 \left[%
-|ijk\rangle-|jik\rangle+|jki\rangle+|kji\rangle\right] , \label{MS2} \\
|MS3\rangle &=& \frac 1{\sqrt{1-q^2-q+q^3}}\frac 12 \left[%
|ijk\rangle-|jik\rangle-|jki\rangle+|kji\rangle\right] , \label{MS3} \\
|MS4\rangle &=& \frac 1{\sqrt{1-q^2-q+q^3}} \frac 1{\sqrt{12}}\left[%
|ijk\rangle+|jik\rangle-2|ikj\rangle+|jki\rangle -2|kij\rangle+|kji\rangle%
\right] ,  \label{MS4}
\end{eqnarray}
where $i,j,k=1,2,3$ . Also, the cases $i=j$, $i=k$, $j=k$ and
$i=j=k$ are automatically included in the expressions of
Eqs.~(\ref{S}) to (\ref{MS4}) up to a normalization factor which
is $q$-independent. Evidently, the above basis states can be built
from the well known procedure based on the Young tableaux
method~\cite{Greiner}, or through the diagonalization of the
overlap
matrix obtained from all possible order permutations from the state $%
A_i^{\dagger }\,A_j^{\dagger }\,A_k^{\dagger }|0\rangle$ . In fact, the $q$%
-polynomials which appear in the square roots in Eqs.~(\ref{S}) to (\ref{MS4}%
), correspond to the eigenvalues of the overlap matrix and measure the
degree of violation of statistics in the system.

Finally, we remark that for the quon algebra one may also define the
transition operator $N_{ij}$, obeying the commutation relations\cite
{GreenPRD,Melja}
\begin{equation}
[N_{ij},A^{\dag}_k ] = {\delta}_{jk} A^{\dag}_i, \hspace{1.0cm}[N_{ji},A_k
]=-{\delta}_{jk} A_i .
\end{equation}
The $N_{ij}$ operator is given by an infinite series expansion in terms of
the quonic annihilation and creation operators, and its first few terms are
given by
\begin{eqnarray}
N_{ij}=A^{\dag}_i A_j + (1-q^2)^{-1} \sum_{\gamma}
(A^{\dag}_{\gamma}A^{\dag}_{i} -q A^{\dag}_{i}A^{\dag}_{\gamma}) (A_{j}
A_{\gamma} -q A_{\gamma}A_{i} ) + \cdots .
\end{eqnarray}
The number operator $N_i$ is obtained when $i=j$.


\subsection{The quon algebra and composite particles}


If one writes $q=1-x$, the deformed commutator, Eq.~(\ref{qAcomm}), can be
written as
\begin{equation}
[A_{\alpha}, A^{\dag}_{\beta}]=\delta_{\alpha \beta} -
xA^{\dag}_{\alpha}A_{\beta}\;.  \label{xdef}
\end{equation}
The similarity of this relation with Eq.~(\ref{Acomm}) is evident. In some
sense, the product of fermion operators $a^{\dag}_{\mu^{\prime}}a_{\mu}$ and
$b^{\dag}_{\mu}b_{\nu^{\prime}}$ weighted by the $\Phi$'s in Eq.~(\ref{Acomm}%
) is effectively modelled by the term $xA^{\dag}_{\alpha}A_{\beta}$ in Eq.~(%
\ref{xdef}).

We note that the commutation relations of Eq.~(\ref{AAcomm}) between $%
A_{\alpha}$,$A_{\beta}$ and $A^{\dag}_{\alpha}$,$A^{\dag}_{\beta}$ are not
algebraic relations valid for quons. Since we wish to describe a system of
``identical'' composite bosons, we impose that the state vectors of a
many-body composite particles system are invariant by the permutation of the
particle indexes. So we assume that the physical subspace which is adequate
for the description of a bosonic composite system is composed only of
totally symmetric states. To build the physical subspace we have to project
out from the basis of states only totally symmetric states.

It is possible to show that the most general symmetric state for a system of
$N$ quons can be written as~\cite{SidRic}:
\begin{equation}
|n_\alpha n_\beta n_\gamma ...;S\rangle=\sqrt{\frac{n_\alpha !n_\beta
!n_\gamma !...} {N![N]!}}{\widehat{S}}_N(A_{\alpha}^{\dagger })^{n_\alpha
}(A_{\beta}^{\dagger })^{n_{\beta}}(A_{\gamma}^{\dagger
})^{n_{\gamma}}...|0\rangle , \label{NSym}
\end{equation}
where ${\widehat{S}}_N$ is an operator that generates all possible
combinations that are symmetric under the permutation of any of the creation
operators, $n_\alpha +n_\beta +n_\gamma +...=N$ and~\cite{Kibler,chai}
\begin{equation}
\lbrack N]=\frac{1-q^N}{1-q},  \label{caixote}
\end{equation}
with $[N]!=[N][N-1]....[2][1]$ and [0]!=1. Another important result that we
are going to use next is the following~\cite{SidRic}:
\begin{equation}
A_\alpha |n_\alpha n_\beta n_\gamma ...;S\rangle=\sqrt{\frac{[N]}N}\sqrt{%
n_\alpha }|n_\alpha -1,n_\beta n_\gamma ....;S\rangle . \label{aNSym}
\end{equation}

This last expression allows one to calculate matrix elements between
symmetric quonic states with any number of quons. Using Eq.~(\ref{aNSym})
one can easily show that, within the physical subspace, i.e., the one built
only from totally symmetric states, $|\psi,S\rangle$, the commutation
relations of Eq.~(\ref{AAcomm}) are now valid also for the quons
\begin{eqnarray}
\langle\psi,S|\left(A^{\dag}_{\alpha} A^{\dag}_{\beta}- A^{\dag}_{\beta}
A^{\dag}_{\alpha} \right)|\psi,S\rangle=0~,~
\end{eqnarray}
\begin{eqnarray}
\langle\psi,S|\left(A_{\alpha} A_{\beta}- A_{\beta} A_{\alpha}\right)
|\psi,S\rangle=0~.
\end{eqnarray}
In this way, the analogy between a composite particle and a quon is complete.


\subsection{Free composite gas}
\label{subsec:free}

Now let us consider a system of $N$ composite bosons in a large box of
volume $V$ at zero temperature. If the bosons were ideal point-like particles,
the ground state of the system would be the one where all bosons condense in the
zero momentum state. In the case of composite bosons, the closest analog of
the ideal gas ground state is

\begin{equation}
|N\rangle={\frac{1}{\sqrt{N!}}}(A^{\dag}_0)^N|0\rangle\;,  \label{Nstate}
\end{equation}
where $A^{\dag}_0$ is the creation operator of a composite boson in its
ground state (ground state $\Phi$) and with zero center of mass momentum.
Due to the composite nature of the bosons, this state incorporates
kinematical correlations implied by the Pauli exclusion principle which
operates on the constituent fermions. Among other effects, the Pauli
principle forbids the macroscopic occupation of the zero momentum state. The
closest analog to the boson occupation number in the state of Eq.~(\ref
{Nstate}) is
\begin{equation}
N_0={\frac{\langle N|A^{\dag}_0A_0|N\rangle}{\langle N|N\rangle}}\;.
\label{number}
\end{equation}

In order to evaluate Eq.~(\ref{number}), we consider a spin zero boson and
use for the spatial part of $\Phi$ a simple Gaussian form such that the
r.m.s radius of the boson is $r_0$. To lowest order in the density of the
system $n=N/V$, $N_0$ is given by~\cite{us}
\begin{equation}
N_0=N\left(1 - \gamma n r_0^3\right)\;,  \label{N0}
\end{equation}
where $\gamma$ is a geometrical factor. The exact value of the geometrical
factor $\gamma$ depends on the form of the internal wave-function of the
composite boson. Depending on the exact form used for the internal boson
wave-function this factor runs from 50 to ~250~\cite{Comb-epl}. Note that
Eq.~(\ref{N0}) is valid only for $\gamma n r_0^3 \ll 1$, i.e. for extremely
low-density systems. For higher-density systems, a nonlinear $N$ dependence
of $N_0/N$ is obtained~\cite{Comb-epl}.

It is apparent from Eq.~(\ref{N0}) that in the limit of infinite tight
binding, $r_0 \rightarrow 0$, one has the familiar Bose-Einstein
condensation. For finite values of the size of the bosons, the effects of
the Pauli principle become important and one has a depletion on the amount
of condensed bosons. Moreover, from Eq.~(\ref{N0}) one has that if the size
of the bound state is of the order of the mean separation of the bosons in
the medium, $d \sim n^{-1/3}$, the depletion is almost total. The depletion
of the condensation is a direct consequence of the deformation of the boson
algebra by the term $\Delta_{\alpha \beta}$.

In the same spirit as in the previous case, we take as the closest analog of
the ideal boson gas ground state the $N$ quon state

\begin{equation}
|N\rangle={\frac{1}{\sqrt{[N]!}}}(A^{\dag}_0)^N|0\rangle\;.  \label{qNstate}
\end{equation}
Note that we are using the same symbols for both the composite and
quon algebra annihilation/creation operators.

As before, the operator $A^{\dag}_0A_0$ is the number operator in the zero
deformation limit only. The effect of the deformation can be evaluated
taking the expectation value of $A^{\dag}_0A_0$ in the state $|N\rangle$,
Eq.~(\ref{qNstate}). To evaluate the expectation value, we make use of the
result given by equation Eq.~(\ref{caixote})
\begin{equation}
N_{0}=\langle N|A^{\dag}_{0}A_{0}|N\rangle=[N]~=~ {\frac{ {1-q^{N}}}{{1-q} }}%
\;.  \label{AA}
\end{equation}
Using the above result one obtains to lowest order in $x$,
\begin{equation}
N_0={\frac{1}{x}}[1-(1-x)^N]\simeq N\left(1-{\frac{1}{2}}Nx\right)\;.
\label{xN0}
\end{equation}
Comparing this result with the one of Eq.~(\ref{N0}) it is clear that the
effect of the deformation, for sufficiently low densities, is such that
\begin{equation}
N x \sim 2 \gamma {\frac{N r_0^3 }{V}}\,.
\label{x}
\end{equation}
That is, the effect of the deformation parameter is proportional to the
ratio of the volume occupied by the bosons to the volume of the system.
Note that for densities not extremely low, a nonlinear $N$ dependence is
expected for the deformation parameter, as remarked above.

\section{The Gross-Pitaeviskii Equation for Composite Bosons}
\label{sec:qGP}

We now use the quon algebra formalism in the BEC. We consider a system of $N$
composite bosons interacting in a spherical harmonic oscillator trap. We
assume that the effective Hamiltonian describing such system is given by
\begin{eqnarray}
H &=& T+V+V_{trap}  \nonumber \\
&=& \sum_{\alpha,\beta} \langle \alpha |T|\beta \rangle A^{\dag}_{\alpha}
A_{\beta} ~+~ {\frac{1}{2}} \sum_{\alpha,\beta,\gamma,\delta} \langle \alpha
\beta |V| \gamma \delta \rangle A^{\dag}_{\alpha} A^{\dag}_{\beta}
A_{\gamma} A_{\delta} + \sum_{\alpha,\beta} \langle \alpha |V_{trap}| \beta
\rangle N_{\alpha \beta},
\end{eqnarray}
where $T$, $V_{trap}$ and $V$ correspond to the kinetic energy, trap
harmonic oscillator potential and the interaction among the composite bosons
respectively. We take as usual
\begin{eqnarray}
T(\vec{x})=-{\frac{\hbar^2 }{{2m}}} {\triangle}_{\vec{x}} ,
\end{eqnarray}
and
\begin{eqnarray}
V(\vec{x},\vec{y})=g {\delta}(\vec{x}-\vec{y}) .
\end{eqnarray}
To obtain the equation describing the condensate of composite bosons we
assume that the lowest energy state is given by
\begin{eqnarray}
|\psi\rangle = {\frac{{(A^{\dag}_0 )^{N}}}{\sqrt{[N]!}}} |0\rangle .
\end{eqnarray}
So we impose the variational principle
\begin{eqnarray}
\langle \psi ,S |H| \delta \psi ,S \rangle =0~,  \label{VARPri}
\end{eqnarray}
where the arbitrary variational symmetrized state has the form
\begin{equation}
|\delta \psi, S\rangle = \hat{S} A^{\dag}_{\mu} A_0 |\psi,S \rangle~,~ (\mu
\neq 0)
\end{equation}
and the symbol $\hat{S}$ is a symmetrizer operator. Eq.~(\ref{VARPri}) is
equivalent to the expression
\begin{eqnarray}
\langle \psi ,S | \sum_{\alpha \beta} \left( \langle \alpha |T|\beta \rangle
A^{\dag}_{\alpha} A_{\beta} + {\frac{1}{2}} \sum_{\gamma \delta } \langle
\alpha \beta |V| \gamma \delta \rangle A^{\dag}_{\alpha} A^{\dag}_{\beta}
A_{\gamma} A_{\delta}+ \langle \alpha |V^{trap}| \beta \rangle N_{\alpha
\beta} \right)| \delta \psi ,S \rangle = 0 .  \label{VarEq}
\end{eqnarray}
From Eq.~(\ref{aNSym}) one obtains the matrix elements
\begin{eqnarray}
\langle \psi ,S | A^{\dag}_{\alpha} A_{\beta} |\delta \psi ,S \rangle &=& {%
\frac{[N]}{N}} \sqrt{N} \delta_{\alpha 0} \delta_{\beta \mu} , \\
\langle \psi ,S | N_{\alpha \beta} |\delta \psi ,S \rangle &=& \sqrt{N}
\delta_{\alpha 0} \delta_{\beta \mu} \langle \psi ,S | A^{\dag}_{\alpha}
A^{\dag}_{\beta} A_{\gamma} A_{\delta} |\delta \psi ,S \rangle  \nonumber \\
&=& {\frac{[N]}{N}} {\frac{[N-1]}{{N-1}}} \sqrt{N} (N-1) ( \delta_{\alpha 0}
\delta_{\beta 0} \delta_{\gamma \mu} \delta_{\delta 0} + \delta_{\alpha 0}
\delta_{\beta 0} \delta_{\gamma 0} \delta_{\delta \mu} ).
\end{eqnarray}
Using these in Eq.~(\ref{VarEq}), one obtains
\begin{eqnarray}
{\frac{[N]}{N}} T_{0 \mu} + V^{trap}_{0 \mu} + {\frac{{[N][N-1]}}{N}}
V_{000\mu} = 0 .
\end{eqnarray}
Therefore, the variational principle leads in coordinate space to the
equation
\begin{eqnarray}
\left[ -{\frac{\hbar^2 }{{2m}}} {\triangle}_{\vec{x}} + {\frac{N}{{[N]}}}
~V^{trap}(\vec{x}) + g [N-1] |\phi(\vec{x})|^2 \right] \phi(\vec{x}) = \epsilon
\phi(\vec{x}).  \label{qGP}
\end{eqnarray}
This equation may be interpreted as a generalization of the
Gross-Pitaeviskii equation to quons and, as mentioned previously will be
denoted as qGP.

\section{The Solution of the Quon Gross-Pitaeviskii Equation and
Applications}
\label{sec:numer}

Many authors have recently presented different methods to solve the usual GP
equation. Here, we follow a variational approach, in which we expand the
ground state wave-function in a three-dimensional harmonic oscillator basis
with angular momentum $l=0$. Good convergence has been achieved within the
method and we were able to reproduce to very good accuracy results
obtained in the literature with other methods when we go to the limit $q=1$.

Initially we consider the case of an attractive interaction, represented by a
negative value of the scattering length \cite{Bec}. Of particular interest
is the study of the well known collapse, which occurs when the number of
atoms in the trap reaches a critical number for which the system becomes
unstable. In that case, the behavior of the GP solution is characterized by
the dimensionless constant
\begin{equation}
k=\frac{N_{0}|a|}{b_{t}},  \label{colapse}
\end{equation}
where $b_{t}$ is the trap oscillator length and $a$ the $s$-wave scattering
length. Several authors (see for instance Ref.~\cite{Roberts} and references
therein) have calculated the critical value ($k_{c}$) for this constant
using different methods. The value, for an isotropic trap, was always very
close to $k_{c}=0.575$. Recent measurements~\cite{Roberts} indicate a value
for $k_c$ that is between $\sim 10\%$ and $25\%$ smaller than the theoretical
predictions.

Let us now consider now the qGP equation. According to our discussion in
Section~(\ref{sec:Comp-q}), one may interpret the number of condensed atoms
as $N_{0}=[N]$. Eq~(\ref{qGP}) can be rewritten in terms of
dimensionless variables in the same fashion as done in the normal GP
equation~\cite{Bec}:
\begin{eqnarray}
\left[ - {\triangle}_{\vec{\widetilde{x}}} + \widetilde{r}^{2} - 8\pi k(q) |{%
\phi}(\vec{\widetilde{x}})|^2 \right] {\phi}(\vec{\widetilde{x}}) = 2 \,%
\widetilde{\epsilon} \, \phi(\vec{\widetilde{x}}) .  \label{res-qGP}
\end{eqnarray}
Here we have used explicitly the oscillator form for the potential and have
written the interaction constant in terms of the scattering length $a$ as
$g=4\pi \hbar^{2}a/m$. The tilde means energy in units of the oscillator
energy, $\widetilde{\epsilon} ={\epsilon}/\hbar\omega_{t}$, and length is
expressed in units of the modified oscillator constant ,
${\widetilde{x}}=x/b_{t}(q)$, with
\begin{equation}
b_t(q) = b_t\left(\frac{[N]}{N}\right)^{1/4} .
\end{equation}
In addition,
\begin{eqnarray}
k(q)={\frac{|a|[N]}{b_{t}}} {\frac{[N-1]}{[N]}}\left({\frac{N}{[N]}}%
\right)^{1/4} \simeq k \left({\frac{N}{[N]}}\right)^{1/4}.
\end{eqnarray}

Obviously the qGP equation of Eq.~(\ref{res-qGP}) leads to a critical value
$k_c(q) = 0.575$. However, in the present case the critical constant to be
compared with the experimental value is given by
\begin{equation}
k_{c} = 0.575 \left(\frac{[N]}{N}\right)^{1/4} .
\end{equation}
For $N_{0}=[N]= 2000$ and using $x$ independent of $N$ and as small as
$x = 3\times 10^{-4}$ (or $q = 1 - x = 0.9999$), one finds that it is
possible to fit the experimental result $k_{c}=0.522$~\cite{Roberts}.
This value of $x$ should be compared to the expression coming from our
estimate of the deformation parameter in Eq.~(\ref{x}). Using for $r_{0}$
the scattering length $|a|\sim 15 \AA$, and using the linear dimension of
the trap as given by $b_{t}\sim 100\AA$~\cite{Cornish} one obtains precisely
\begin{equation}
x = 2\gamma\frac{|a|^3}{ b^3_t} \sim 10^{-4},  \label{exper-x}
\end{equation}
where for the geometrical factor $\gamma$ we used the value $\gamma=100$,
as estimated in Ref.~\cite{Comb-epl}. This result is of the correct order
of magnitude to bring the theoretical value for the critical constant within
the range of experimental values.

As a second application, we consider the case of a repulsive
interaction. In this case the number of atoms in the trap
can reach large numbers (N up to $\sim 10^{7}$), which can raise
the question of validity for the use of the GP equation for such
high densities. In a recent calculation \cite {fabrocini}, effects
that go beyond the usual GP (or mean-field) solution were also
considered and shown to lead to a systematic, small increase
in the chemical potential of the condensate. In our formalism we may
obtain the same behavior by preserving the original GP dynamics but
relaxing the condition that the particles are true bosons. To see how
this happens , we now rewrite Eq.~(\ref{qGP}) in a slightly different
form:
\begin{eqnarray}
\left[ H_{osc} + \left({\frac{N}{{[N]}}}-1\right) V^{trap}(\vec{x})
+ g [N-1] |\phi(\vec{x})|^2 \right] \phi(\vec{x}) = \epsilon \, \phi(\vec{x}) ,
\end{eqnarray}
where $H_{osc}$ is the usual three-dimensional harmonic oscillator
Hamiltonian. Following our previous interpretation that the number of condensed
atoms is $[N]$, we conclude that only the term proportional to
$N/[N]-1$ differs from the usual GP equation. However, for $q$ close enough to
$1$, this term is small ($\ll 1$) even for large values of $N$, and therefore
one may treat this term as a perturbation to the non-deformed solution.
At this point it is worthwhile to note that this term
will always increase the energy of the condensate by a small
amount. Of course, the increase will depends on the value of $q$ (or $x$)
used. In order to make a phenomenological estimation of the effect of this
perturbation we relate again our results of section~\ref{subsec:free} for
the depletion of the condensate to the one presented in Section~III of
Ref.~\cite{Bec}. There, an estimation for the depletion is given by:
\begin{eqnarray}
\frac{N-N_{0}}{N_{0}}=\frac{5\sqrt{\pi}}{8}\sqrt{a^{3}n(0)} ,
\end{eqnarray}
where $n(0)$ is the central ($r=0$) density of the
condensate. Equating our result, Eq.~(\ref{xN0}), with the above
and taking in to account that $x\ll 1$, we obtain:
\begin{eqnarray}
x\cong \frac{10\sqrt{\pi}}{8N}\sqrt{a^{3}n(0)} .
\label{xx}
\end{eqnarray}

Note that a dependence of the deformation parameter on the number of atoms
(or on the density) appears very naturally here. As we have discussed in
Section~\ref{subsec:free}, an $N$-independent deformation parameter $x$
is physically reasonable only for low density systems. In the present
situation of a repulsive interaction, the number of atoms in the trap
can be many orders of magnitude larger than in the attractive case and
it is gratifying to see the internal consistency of our formalism, in
that it is capable of describing both extremes in a very natural way.

In order to perform numerical calculations, one needs $n(0)$, which can
be taken from the $x=0$ solution of the GP equation and the value for $N$,
which in turns depends on $x$. However, for our purposes here we may
evaluate the order of magnitude of $x$ making $N=N_{0}$ in Eq.~(\ref{xx}).
We also use $a/b_{t}=5\times 10^{-2}$. This last value might seem quite big
when compared to the one used for example in Refs.~\cite{Bec} and
\cite{Roberts}. However, recent experimental developments~\cite{Cornish}
allow to change the scattering length by orders of magnitude thanks
to a technique known as Feshbach resonance. The value chosen here is consistent
with, for example, the experiment reported in Ref.~\cite{Cornish}.
In Table~I we present our numerical results for the chemical
potential $\epsilon $, for different $N_{0}$ values. Note that,
although our initial conditions are different from the ones chosen
for the calculations in Ref.~\cite{fabrocini}, the results are
qualitatively equivalent, in the sense that we get always a
bigger value for the chemical potential, as compared to the usual
GP solution.

\section{Conclusions}

\label{sec:concl}

We have considered in this work the quon algebra to describe in an
effective and phenomenological way the departure from a purely bosonic
behavior of a system of composite bosons. The formalism was developed
previously and relies on the projection of the whole quonic space onto the
symmetric subspace. The main idea is to preserve the bosonic behavior
and leave to the deformation parameter the description of possible
deviations. As a specific application we have considered the derivation
of the Gross-Pitaeviskii equation within the quon algebra, which can be
done in a straightforward way using our formalism. The interpretation of
the modified Gross-Pitaeviskii equation obtained is consistent with our
initial qualitative considerations.

As for numerical calculations we have in first place solved our
deformed GP equation for the collapse in the case of an attractive
force in a trapped bosonic gas. Reinterpreting the value of the
critical constant for the collapse, we obtained that the
deformation of the bosonic algebra changes the value of this
constant, and the change is in the right direction to explain
experimental results. Of course, the modification depends on the
chosen value for the deformation parameter $q$, which is not easy
to estimate at the conditions of the experiment. We have found
that a decrease of order $10^{-4}$ compared to the bosonic limit
$q=1$, and independent of the number of atoms in the trap, brings
the critical value within the recent experimental measurements.
Secondly, we have calculated the chemical potential for the case
of a repulsive force between the trapped atoms. This is a more
favorable situation to test our model, since recently it became
experimentally feasible to create condensates with a much larger
number of atoms as compared to the attractive case. Using a
phenomenological estimation for the deformation parameter, for
which a density dependence arises very naturally, we are able to
calculate the correction in the chemical potential.

Natural extensions of the present work would include the
investigation of the effect of the $q$-deformation on other
observables of the BE condensates. Also, there are many other
interesting systems for which the quon algebra could be useful to
describe deviations from a purely bosonic behavior of many-body
systems of composite bosons. One example, which seems
particularly interesting in the present context, is the case of
excitons~\cite{Comb-epl}, which are boson-like particles formed by
a electron-hole pair in a semiconductor. Work in these directions
is underway.

\begin{acknowledgments}

This work was partially supported by  CNPq.

\end{acknowledgments}

\newpage
\begin{table}
\caption{Chemical potential $\epsilon$ - in units of $\hbar\omega_{t}$ -
of the condensate for four values of the  number of atoms $N_{0}$, using
the usual GP formalism and the qGP equation. The values of $x$ used in
each case are also shown and were obtained as explained in the text.}
\begin{ruledtabular}
\begin{tabular}{cccc}
$N_{0}$ & $\epsilon_{\rm GP}$ & $\epsilon_{\rm qGP}$ & $x$ \\ \hline
1000 & 7.31 & 7.47 & $1.0\times10^{-3}$ \\
10000 & 17.84 & 18.32 & $1.3\times10^{-4}$ \\
100000 & 44.59 & 46.69 & $2.0\times10^{-5}$ \\
1000000 & 111.94 & 131.16 & $5.1\times10^{-6}$ \\
\end{tabular}
\label{table}
\end{ruledtabular}
\end{table}

\end{document}